\documentclass[aps,prl,twocolumn,showpacs,floatfix,superscriptaddress,nofootinbib]{revtex4-2}
\usepackage{graphicx}%
\usepackage{dcolumn}%
\usepackage{bm}%
\usepackage{physics}
\usepackage{hyperref}%
\usepackage{color}
\usepackage{amsmath}
\usepackage{amssymb}
\usepackage{ulem} %
\usepackage{multirow}
\usepackage{booktabs}
\usepackage{blkarray}

\newcommand{\scarZ}[1]{\mathbb{Z}_{#1}}
\newcommand{\Ztwo}{\mathbb{Z}_2}
\newcommand{\Zthree}{\mathbb{Z}_3}
\newcommand{\Zfour}{\mathbb{Z}_4}
\newcommand{\Zfive}{\mathbb{Z}_5}
\newcommand{\Zzero}{\mathbb{Z}_0}
\newcommand{\Ntrash}{N_{\text{trash}}}
\newcommand{\Nshots}{N_{\text{shots}}}
\newcommand{\Niter}{N_{\text{iter}}}

\def\sx#1{\sigma^{\rm x}_{#1}}
\def\sy#1{\sigma^{\rm y}_{#1}}
\def\sz#1{\sigma^{\rm z}_{#1}}
\def\tx#1{\tau^{\rm x}_{#1}}
\def\ty#1{\tau^{\rm y}_{#1}}
\def\tz#1{\tau^{\rm z}_{#1}}

\definecolor{gr4} {RGB}{34,198,34}

\begin{document}
\title{Unsupervised detection of decoupled subspaces: many-body scars and beyond}%
\author{Tomasz Szo\l{}dra} 
\affiliation{Instytut Fizyki Teoretycznej, 
Uniwersytet Jagiello\'nski,  \L{}ojasiewicza 11, PL-30-348 Krak\'ow, Poland}
\author{Piotr Sierant} 
\affiliation{ICFO-Institut de Ci\`encies Fot\`oniques, The Barcelona Institute of Science and Technology, Av. Carl Friedrich
Gauss 3, 08860 Castelldefels (Barcelona), Spain}
\author{Maciej Lewenstein} 
\affiliation{ICFO-Institut de Ci\`encies Fot\`oniques, The Barcelona Institute of Science and Technology, Av. Carl Friedrich
Gauss 3, 08860 Castelldefels (Barcelona), Spain}
\affiliation{ICREA, Passeig Lluis Companys 23, 08010 Barcelona, Spain}
\author{Jakub Zakrzewski} 
\affiliation{Instytut Fizyki Teoretycznej, 
Uniwersytet Jagiello\'nski,  \L{}ojasiewicza 11, PL-30-348 Krak\'ow, Poland}
\affiliation{Mark Kac Complex Systems Research Center, Uniwersytet Jagiello{\'n}ski, Krak{\'o}w, Poland}

\date{\today}

\begin{abstract}

Highly excited eigenstates of quantum many-body systems are typically featureless thermal states. Some systems, however, possess a small number of special, low-entanglement eigenstates known as quantum scars.  We introduce a quantum-inspired machine learning platform based on a Quantum Variational Autoencoder {(QVAE)} that detects families of scar states in spectra of many-body systems.  Unlike a classical autoencoder, QVAE performs a parametrized unitary operation, allowing us to compress a single eigenstate into a smaller number of qubits. We demonstrate that the autoencoder trained on a scar state is able to detect the whole family of scar states sharing common features with the input state.
We identify families of quantum many-body scars in the PXP model beyond the $\Ztwo$ and $\Zthree$ families and find dynamically decoupled subspaces in the Hilbert space of disordered, interacting spin ladder model. The possibility of an automatic detection of subspaces of scar states opens new pathways in studies of models with a weak breakdown of ergodicity and fragmented Hilbert spaces.

\end{abstract}
\maketitle

\paragraph*{Introduction.}
Recent progress in noisy, intermediate-scale quantum (\textbf{NISQ}) computers \cite{Preskill18, Wu21, Zhong21} lead to a fast development of algorithms suited for use on these machines \cite{Bharti22} with the purpose of achieving a quantum advantage in various areas: physics, machine learning, quantum chemistry, and combinatorial optimization. Of particular importance are variational quantum algorithms \cite{Cerezo21rev}, in which quantum circuits are applied to quantum states, whose parameters are optimized with classical feedback loops. Physical applications include variational quantum eigensolvers \cite{Peruzzo14, McClean16, Jones19}, algorithms for ground state preparation \cite{Ho19a}, time evolution simulations \cite{Li17var, Yuan19, Endo20} or quantum variational autoencoders (\textbf{QVAE}) \cite{Romero17, Bravoprieto21, Cao21}.
The autoencoders encode the input data into a reduced representation and then use it to reconstruct the data with the optimal fidelity. As such, autoencoders are basic tools for data compression in machine learning. In turn, the task of QVAE is to realize the unitary transformation that transfers the input entangled n-qubits state
into a product state of $k$, relevant, entangled qubits and $n-k$ ``trash'' separable qubits (see Eq.~\ref{eq:QVAE} below). 
 QVAE have been realized experimentally in a photonic device \cite{Pepper19} and recently employed in investigation of quantum phase transitions \cite{Kottmann21c}. In this work we demonstrate the applicability of QVAE in an analysis of properties of highly excited eigenstates in quantum many-body systems.

According to Eigenstate Thermalization Hypothesis (\textbf{ETH}) \cite{Deutsch91, Srednicki94, Alessio16}, a small subsystem of an isolated, interacting quantum many-body system is described by a thermal density matrix after a long time evolution, irrespectively of the initial non-equilibrium state. 
However, some systems violate this paradigm of quantum ergodicity and exhibit a long-time behavior dependent on the initial state. Examples of such non-ergodic systems include integrable systems \cite{Sutherland04} and many-body localized phases in the presence of quenched disorder \cite{Gornyi05,Basko06, Nandkishore15, Alet18, Abanin19}. 
Another mechanism of ergodicity breaking in a form of persistent oscillations for particularly chosen initial states has been discovered in the experiment with ultracold Rydberg atoms \cite{Bernien17}. This behavior arises due to the presence of few atypical, almost equally spaced eigenstates with low entanglement entropy, the so-called quantum many-body scars (\textbf{QMBS}) \cite{Turner18, Turner18q} that are embedded in the otherwise thermal spectrum of a quantum many-body system. For initial states with high overlap with a few QMBS, one observes long-lived oscillations of observables, whereas for generic initial conditions the system quickly approaches the thermal equilibrium state. Several theories explaining the emergence of QMBS were proposed starting long time ago with the notion of ``scars of symmetries'' \cite{Delande87} (see also \cite{Serbyn20}): a spectrum generating algebra \cite{Moudgalya18en, Iadecola20}, Krylov restricted thermalization \cite{Moudgalya21}, projector embedding \cite{Shiraishi17} and the presence of symmetric, coupled subspaces \cite{Turner21}. The QMBS occur in PXP model \cite{Lesanovsky12}, describing Rydberg atoms chain, but also in AKLT model \cite{Moudgalya18, Mark20}, quantum local random networks \cite{Surace21q}, frustrated magnetic lattices \cite{Lee20}, lattice gauge theories \cite{Surace20, Magnifico20, Aramthottil22}, optical lattices \cite{Zhao20} or spin systems \cite{Iadecola19, Iadecola19a}.

The aim of this work is to provide a scheme to detect families of QMBS based on QVAE. A \textit{family} of QMBS is formed by eigenstates which: \textit{a}) have an increased overlap with some 'parent' state, \textit{b}) are characterized by a sub-volume-law entanglement entropy. The property \textit{a}) enables our algorithm to implicitly extract the features of the parent state from a single eigenstate in the training and encode them in QVAE. The probability that training succeeds is enhanced by the property \textit{b}). The performance of the trained autoencoder on other eigenstates serves as a measure of similarity between the eigenstates.
The other representatives of the family of QMBS are found as eigenstates for which the performance of QVAE is significantly better than the typical behavior. In the following, we first describe details of the scheme. Then we apply it to detect the $\Ztwo$ family and to discover new families of QMBS in the PXP model \cite{Lesanovsky12, Turner18} and to identify subspaces of decoupled eigenstates in the spin ladder of \cite{Iadecola19}.

\paragraph*{Quantum variational autoencoder. }
The QVAE aims to compress the $n$-qubit input state $\ket{\psi}$ into a $k$-qubit state $\ket{\phi}$ (where $k<n$), i.e., to perform a unitary transformation $U(\bm{\theta})$ parametrized by the circuit parameters $\bm{\theta}$,
\begin{equation}
        \ket{\psi} \rightarrow U(\bm{\theta})\ket{\psi} = \ket{\phi}\otimes \ket{0}^{\otimes (n-k)},
        \label{eq:QVAE}
\end{equation}
where the last $n-k\equiv N_{\text{trash}}$ qubits are called "trash" qubits. The QVAE circuit starts with randomly initialized parameters $\bm{\theta}$ which are variationally adjusted (here - using Simultaneous Perturbation Stochastic Approximation optimizer \cite{Spall97, Spall98, Kandala17, Qiskit21short}) to "optimal" value $\bm{\theta}=\bm{\theta^*}$ satisfying Eq. (1) for a given set of input states $\lbrace \ket{\psi_i} \rbrace$. At each optimization step, the unitary $U(\bm\theta)$ is applied on $\lbrace \ket{\psi_i} \rbrace$ and the trash qubits are measured giving either '0' or '1'. The total number of '1' defines the cost function that the optimizer tries to minimize in the next parameter update. (Cost function is the Hamming distance between the measured bitstring of '1' and '0' on all trash qubits and the desired $\ket{0}^{\otimes (n-k)}$ trash qubit configuration.)
Unlike for classical autoencoders, training a unitary encoder $U(\bm{\theta})$ automatically provides the decoder $U^\dagger(\bm{\theta})$ that can be applied to the compressed state  $\ket{\phi}\otimes \ket{0}^{\otimes (n-k)}$ to reconstruct the input state $\ket{\psi}$. The cost function fulfills the requirement of locality on the trash qubits which is critical for circumventing the ''barren plateaus'' of the cost landscape and trainability of the model, see \cite{Cerezo21, McClean18} for further details.

The architecture of the quantum circuit has to be expressible (i.e. able to encode a large class of quantum states with a few trainable parameters $\bm{\theta}$) and to possess a large entangling capability to transfer the entanglement of the whole system out of the trash qubits \cite{Sim19}. Building on the previous results \cite{Sim19, Cerezo21} we choose Alternating Layered Ansatz consisting of layers with single-qubit rotations around the $y$ axis by an angle $\theta${$\in[0,2\pi]$}, $R_y(\theta) = \exp(-i \sigma_y \theta/2)$, and two-qubit controlled-Z ($CZ$)  gates that apply a $\sigma_z$ operator on one of the qubits if the other one is in the state $\ket{1}$ and act as an identity if the other qubit is in the state $\ket{0}$. Each of $L$ layers of QVAE consists of $R_y(\theta)$ rotations of all qubits and $CZ$ entangling operations between the neighboring qubits, with the pairs of entangled qubits alternating from layer to layer following a checkerboard pattern {(see \cite{suppl})}.

In our scheme, as an input state $\ket{ \psi}$ to train QVAE we take a single scar state that belongs to a given family of QMBS in a considered many-body system. To identify the other scars from the {same} family, we evaluate the performance of QVAE on eigenstates from the spectrum of the model. The numerical complexity of the procedure is thus $O(\mathcal{D})$ times the number of iterations for QVAE training and $O(\mathcal{D}^2)$ for the comparison of eigenstates where $\mathcal{D}$ is the dimension of the Hilbert space. This is lower than the exact diagonalization cost $O(\mathcal{D}^3)$ required for generation of the input data.

\paragraph*{Scars in the PXP model.} 
The PXP Hamiltonian reads
\begin{equation}
        \hat{H} = \sum_i \hat{P}_{i-1} \sigma_i^x \hat{P}_{i+1} 
\end{equation}
with periodic boundary conditions, where the projectors $\hat{P}_i = (1-\sigma_i^z)/2$ ensure that neighboring spin up states are separated at least by one lattice site, hence implementing the Rydberg blockade phenomenon \cite{Browaeys20} as a constraint on the Hilbert space. Certain specific initial states like $\Ztwo=\ket{0101\dots}$, $\Zthree=\ket{001001\dots}$ and product states that contain domain walls between $\Ztwo$ and $\Zthree$ configurations give rise to persistent long-time oscillations of the local observables and the revivals of the wave function, while other states like $\ket{\Zzero} = \ket{0000\dots}$ and $\ket{\Zfour} = \ket{00010001\dots}$ show fast relaxation without revivals. The presence of families of $\scarZ{2}$ and $\scarZ{3}$ quantum scars gives rise to this behavior \cite{ Khemani19, Choi19, Ho19, Iadecola19q, Michailidis20, Moudgalya20, ODea20, Bull20, Surace21}. Some of the scarred states in the PXP model were constructed exactly as MPS with a finite bond dimension \cite{Lin19}, from which the family of $\scarZ{2}$ scars was generated as quasiparticle excitations.

Input data to QVAE corresponds to the eigenstates of the PXP model obtained through exact diagonalization for the system size $N=24$ in the inversion-symmetric and zero-momentum sector with the Hilbert space dimension $\mathcal{D}=2359$ ($D \ll 2^L$ due to constraints and periodic boundary conditions). 
Local constraints of the PXP model allow to reduce the computational cost of the procedure by considering only the projection of QVAE onto the constrained subspace of Hilbert space. To that end, it suffices to substitute $R_y(\theta) \rightarrow \tilde{R}_y(\theta)$, $CZ_i \rightarrow  E_i$ in the circuit, with $\tilde{R}_y(\theta)$ rotating qubit $i$ only if qubits $i-1$, $i+1$ are in the $\ket{0}$ state (identity otherwise), and $E_i$ acting on four qubits $i-1, \dots, i+2$, performing the entangling operation of qubits $i$ and $i+1$ if qubits $i-1, i+2$ are in the state $\ket{0}$ (identity otherwise). Exact matrix forms of these operators are given in \cite{suppl}.
This version of the QVAE will be referred to as the CQVAE. We should note that the translational and inversion symmetry of the original Hamiltonian are not exploited in the CQVAE because these symmetries are manifestly broken by the considered few-qubit gates. Thus, for $N=24$, the CQVAE still acts on $2359$ eigenstates but each of them is expressed in a $103682$-dimensional Hilbert space.

\begin{figure}
	\centering
    \includegraphics[width=\columnwidth]{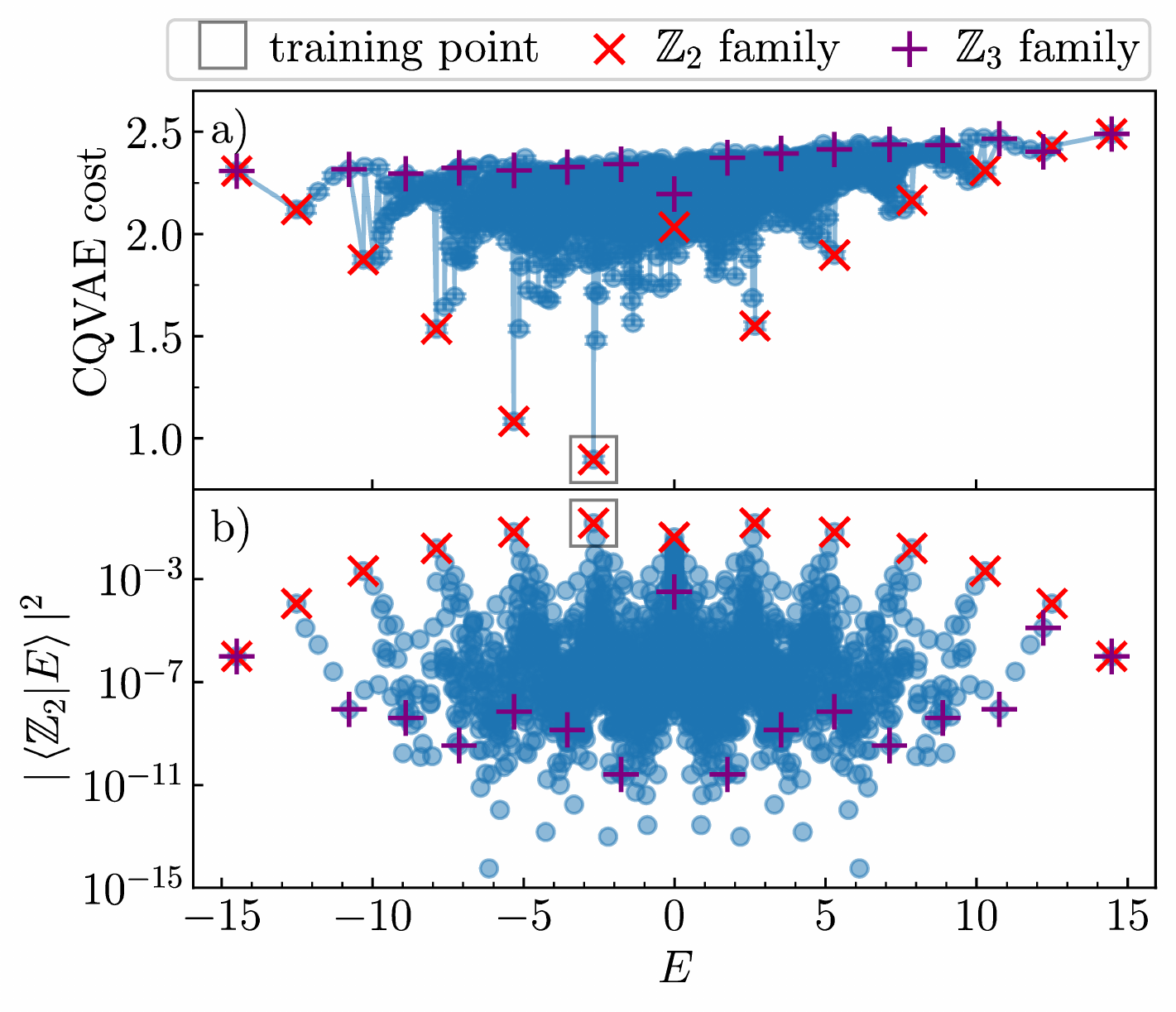}
    \caption{a) Performance of the CQVAE with $\Ntrash=8$ and $L=7$ trained on the strongest $\Ztwo$ scar of the PXP model with $N=24$ close to the middle of the spectrum, applied to all eigenstates. Best performance is observed in the eigenstates from the $\Ztwo$ scars family that have an increased overlap with the $\ket{\Ztwo}$ state, as presented in b). Lines serve as a guide to the eye.  Error bars come from averaging over 32 independent trainings.}
    \label{fig:QVAE_PXP_Z2}
\end{figure}
       
We select a $\Ztwo$ scar at energy $E\approx-2.67$ as the training input state $\ket{\chi_0}$. This state has a significant overlap with the $\scarZ{2}$ configuration, i.e., $|\braket{\chi_0}{\scarZ{2}}|^2 = 0.15$, cf. with a value expected in the high-temperature thermal ensemble $1/\mathcal{D} \approx 0.0004$. Fig.~\ref{fig:QVAE_PXP_Z2}a shows performance of a trained autoencoder on eigenstates from the PXP Hamiltonian spectrum. Indeed, we see that the $\Ztwo$ scars are characterized by a significant drop in the CQVAE cost. Plots of CQVAE reconstruction fidelity (not shown) also reveal high-fidelity peaks on $\Ztwo$ scars. In this way, the family of $\Ztwo$ scars can be identified in an automatic way. Interestingly, the $\Zthree$ family has the largest cost even though it has low entanglement entropy showing that CQVAE learned to distinguish the real space patterns $\Zthree$ scar states from the dominant configurations of $\Ztwo$ family, see FIG.~\ref{fig:QVAE_PXP_Z2}b). Interestingly, at smaller system sizes, e.g., $N=12$, we can recover the "parent" state $\ket{\chi}\approx\ket{\Ztwo}$ for the family by optimizing the \textit{input} of the trained CQVAE to minimize cost, see \cite{suppl} for details.

\begin{figure}
	\centering
	\includegraphics[width=\columnwidth]{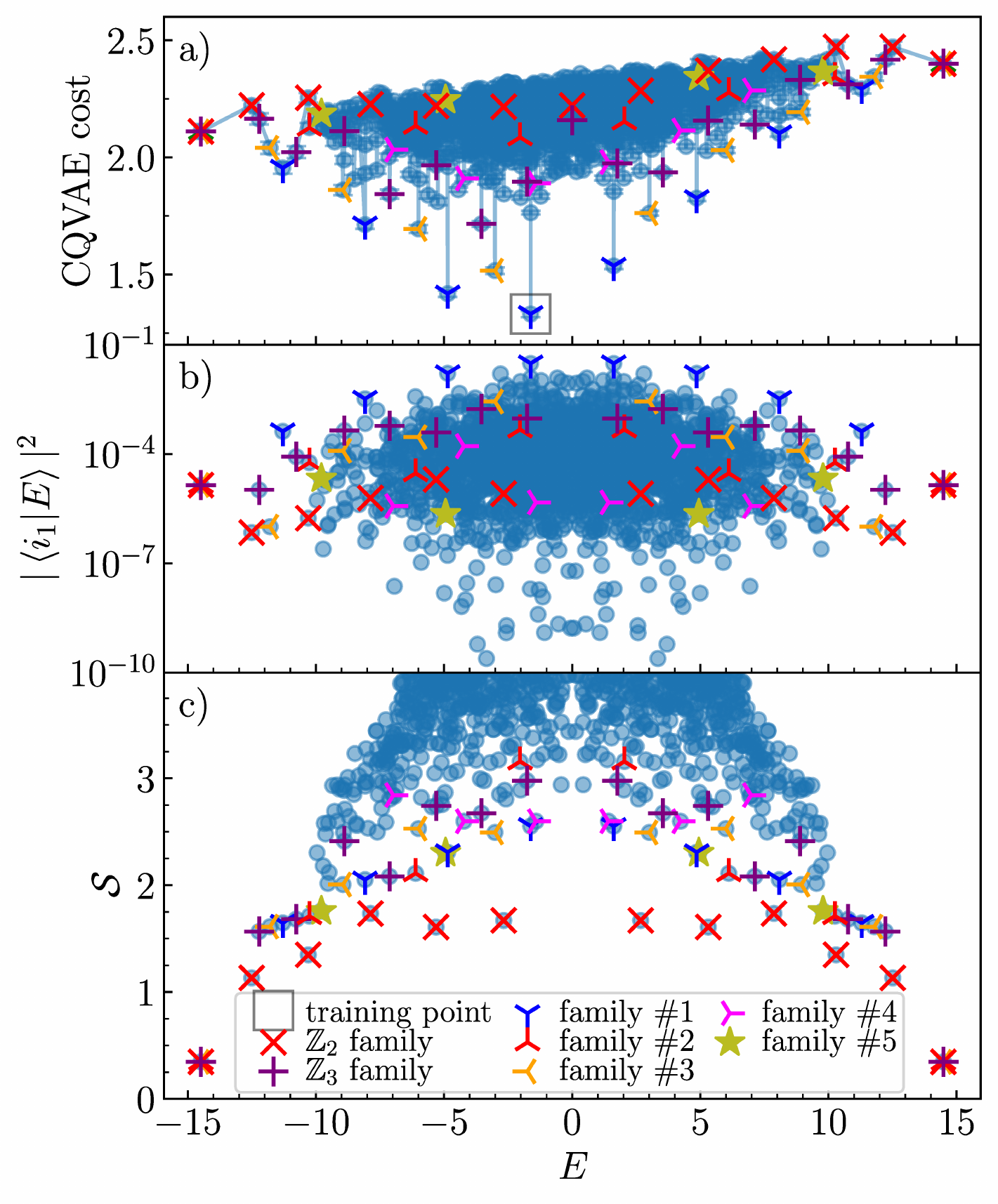}
	\caption{ a) CQVAE cost of eigenstates of the PXP model, $N=24$, reveals a new scar ''family $\#1$'' upon training on one of its representatives. Similar plots for other scarred families found by the algorithm are given in \cite{suppl}. b) Overlaps of the eigenstates with state $\ket{i_1}$. c) Entanglement entropy vs energy.} 
    \label{fig:QVAE_PXP_new_families}
\end{figure}

The next step is to find other scarred families. We select eigenstates with a low entanglement entropy $\mathcal{S} = - \Tr \rho_A \log \rho_A {< 2.7}$, where $\rho_A$ is the reduced density matrix of the half of the spin chain, train the CQVAE on each of them, and calculate the cost on other eigenstates. Pairs of eigenstates that have a low cost when training on both of them are regarded as belonging to the same scar family. This property is transitive, i.e. if eigenstates $\ket{E_1}$, $\ket{E_2}$ have a small cost and $\ket{E_2}$, $\ket{E_3}$ as well, then a set $\ket{E_1}$, $\ket{E_2}$, $\ket{E_3}$ is regarded as one family. Fig.~\ref{fig:QVAE_PXP_new_families} shows 5 new families discovered by the CQVAE in the PXP model for $N=24$, with an example of the training results on one representative of the family $\#1$ (panel a)). The eigenstates from the family $\#1$ are characterized by increased overlaps with several Fock states, the example of $\ket{i_1}=\ket{101010010010100100100100}$ state is shown in FIG.~\ref{fig:QVAE_PXP_new_families}b. Other Fock states with high overlaps with the family $\#1$ contain a mixture of the same number of rearranged three $\Ztwo$ and six $\Zthree$ patterns.  The same holds true for families $\#2$-$\#5$ of scar states found for $N=24$. Patterns with a larger period can be found, e.g., family \#2 has increased overlaps with four $\Zthree$ and three $\Zfour$ configurations. In that way, our QVAE-based scheme allows to explain the presence and identify relations between eigenstates with low entanglement entropy of the PXP model, see Fig.~\ref{fig:QVAE_PXP_new_families}c. Since the new families of scar states do not have a single simple representative Fock state, their classification, especially in system-size independent manner is  more involved -- in \cite{suppl} we show the details of the new families for $N=24$.

\paragraph*{Disordered spin ladder.}
Consider a spin ladder with Hamiltonian 
\begin{subequations}
\label{eq:XXladder}
\begin{align}
\label{eq:XXladdera}
H&=H^{||}+H^{\perp}=\frac{1}{4}\sum_{k=1}^{L-1} h^{||}_{k,k+1}+\frac{1}{4}\sum_{k=1}^{L} h^{\perp}_{k},
\end{align}
where
\begin{align}
\label{eq:XXladderb}
h^{||}_{k,k+1}&=\sx{k} \sx{k+1}+\sy{k}\sy{k+1}+\tx{k}\tx{k+1}+\ty{k}\ty{k+1},\\
h^{\perp}_{k}&=J(\sx{k} \tx{k}+\sy{k} \ty{k})+\Delta_k\, \sz{k} \tz{k}+h_k(\sz{k}+\tz{k}).\nonumber
\end{align}
\end{subequations}
$k=1,\dots,N$ labels the rungs of the ladder, and spins on the left and right legs of the ladder are represented by Pauli matrices $\sigma^\alpha_{k}$ and $\tau^\alpha_{k}$ ($\alpha=\rm x,\rm y,\rm z$), respectively. Values of $h_k$ are drawn from a uniform distribution in the interval $[-h,h]$, and we set  $J=1$, $h=0.1$, $\Delta_k=1$.
The model has a $U(1)$ symmetry associated with the total magnetization $Z=\sum_{k=1}^N(\sigma_k^z + \tau_k^z)$ and a $\Ztwo$ symmetry associated with the exchange of the ladder legs $\sigma^{\alpha}_{k} \leftrightarrow \tau^{\alpha}_{k}$. Even though this model has signatures of quantum ergodicity (e.g. energy levels spacings follow the Wigner-Dyson distribution), one can analytically construct exact invariant subspaces of the Hamiltonian \eqref{eq:XXladder} that are not related to any local conserved quantity as shown in \cite{Iadecola19}. It is first noticed that the eigenstates of $h^\perp_k$ on a single rung are $\ket{S} = (\ket{^0_1} - \ket{^1_0})/\sqrt{2}$ (''singlet''), $\ket{T} = (\ket{^0_1} + \ket{^1_0})/\sqrt{2}$ (''triplet''), $\ket{D} = \ket{^0_0}$ (''doublon''), $\ket{H} = \ket{^1_1}$ (''holon''), where the first (second) row of the vector corresponds to the left (right) leg of the ladder. Product of such states is an eigenstate of the total leg Hamiltonian $H^\perp$. By examining the action of the remaining $H^{||}$ on the two-rung states $\ket{\lbrace ST,SH,TH,HH\dots\rbrace}$ one shows that $H^{||}$ annihilates configurations $ST, TS, HH, DD$, and moves $H$ (or $D$) around if $S$ or $T$ are its neighbours. It follows that configurations $\ket{STSTST\dots}$ and $\ket{TSTSTS\dots}$ are annihilated by $H^{||}$ (they are a ''vacuum background''). Upon inserting a given number of only holons (or doublons) between them, e.g., $\ket{STSH_jTSH_kTS\dots}$, one constructs an invariant subspace with a given number of the four letters that are conserved under the action of the total Hamiltonian $H$. Dimension of such a subspace after $r$ insertions of doublons (or holons) is $\binom{N}{r}$.

\begin{figure}
    \centering
    \includegraphics[width=\columnwidth]{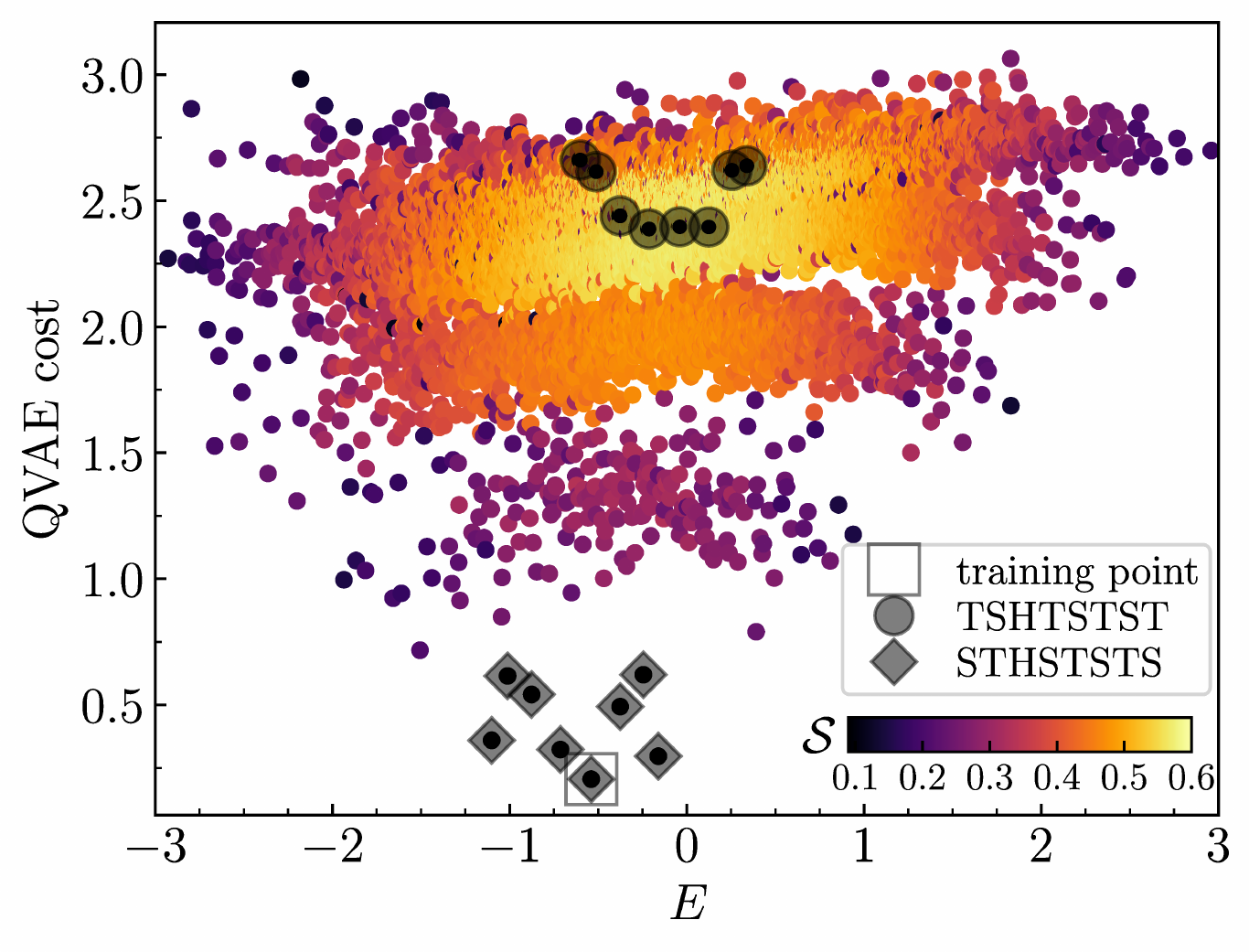}
    \caption{Cost function of QVAE trained on an eigenstate from the 1-holon subspace of the spin ladder model with $N=8$ ($16$ spins in total), evaluated on the eigenstates from $Z=1$ symmetry sector. Colorbar denotes the bipartite entanglement entropy for $4$ rungs. Results averaged over 56 independent trainings of the QVAE with $\Ntrash=5$ and $L=10$.}
    \label{fig:qvae_spin_ladder}
\end{figure}

Construction of invariant subspaces in this model required an involved theoretical insight \cite{Iadecola19}. Here, by employing the QVAE we can detect their presence in an automatic way. Let us restrict only to the $Z=1$ symmetry sector of the Hamiltonian \eqref{eq:XXladder} with $N=8$ rungs and a single disorder realization. We encode the ladder state onto the spatially one-dimensional quantum circuit, by mapping the left (right) leg of the ladder to odd (even) sites of the circuit. In this manner, the neighboring spins of the ladder are mapped to sites of the circuit that are close to each other.
Training of the QVAE on a generic eigenstate and application to all others gives a featureless QVAE cost. However, if we train on an eigenstate from the invariant subspace represented by state $\ket{STSTHSTS}$ with 1 holon $H$, 4 singlets $S$ and 3 triplets $T$, we get a significantly lower error on all 8 eigenstates that span this subspace, see Fig.~\ref{fig:qvae_spin_ladder}. 
The QVAE cost on the other invariant subspace $\ket{TSTSHTST}$ with a different number of singlets and triplets (but an identical entanglement entropy - notice the color scale in Fig. \ref{fig:qvae_spin_ladder}) is comparable to the QVAE cost on generic, highly entangled, eigenstates.
Similarly, a QVAE trained on an eigenstate from the subspace $\ket{TSTSHTST}$ yields small costs on eigenstates from this subspace whereas the cost on eigenstates from $\ket{STSTHSTS}$ is substantial (plot not shown). This is in a full analogy with results for PXP model and demonstrates how QVAE identifies and distinguishes families of scar states.

\paragraph*{Conclusion.} We proposed a scheme based on QVAE that allows to identify families of non-ergodic eigenstates of quantum many-body systems. We  validated our scheme on the $\Ztwo$ family of scar states of the PXP model. Then, our scheme was employed to demonstrate presence of families of scar states in spectrum of PXP model beyond the $\Ztwo$ and $\Zthree$ families. To confirm the generality of our approach, we used it to identify the family of scar states in the disordered spin ladder \eqref{eq:XXladder} as well as in a PXP-like model with a three-body blockade \cite{suppl}. The use of QVAE is crucial in our scheme. QVAE learns from a single, high-dimensional measurement point (a single eigenstate), in contrast to classical autonencoders that require a larger set of training data. The flow of the entanglement entropy through the layers of the autoencoder has a physical meaning and results in a compression of the quantum state. By respecting the laws of quantum mechanics, QVAE becomes a versatile tool in studying eigenstates of many-body systems allowing, for instance, for a direct implementation of the local constraints of PXP model on the QVAE. Finally wee also point out that having the eigenstates one can  try to identify scarred states by calculating the inverse participation ratio (IPR) with a negligible numerical cost instead of costly entanglement entropy. Moreover, instead of exact diagonalization used here, one may use an approximate algorithm, e.g. DMRG-X \cite{Khemani16}, to obtain a subset of eigenstates serving as an input to QVAE.

While all calculations performed here used classical machine, hardware implementation of  {QVAE} on a physical quantum computer seems straightforward \cite{Kottmann21c}. Although preparation of ground states of selected Hamiltonians is possible by the variational quantum eigensolvers, algorithms that provide exited states are more involved \cite{Bharti22}. Hence, the preparation of the input states is the most challenging step of our scheme that is feasible only for limited system sizes. However, to navigate through the exponentially large Hilbert space one can use a prior knowledge about the scar states (e.g. their energy) which may extend the interval of system sizes accessible on current quantum hardware.

\acknowledgements    
We are grateful to Korbinian Kottmann for  valuable suggestions. 
The numerical computations have been possible thanks to   PL-Grid Infrastructure.
This research was partially funded by the Priority Research Area Digiworld under 
the program Excellence Initiative – Research University at the Jagiellonian University in Krakow.
The works of T.S. and J.Z. have been realized within the Opus grant
 2019/35/B/ST2/00034, financed by National Science Centre (Poland). 
  P.S. and M.L. acknowledge support from: ERC AdG NOQIA; Agencia Estatal de Investigación (R\&D project CEX2019-000910-S, funded by MCIN/ AEI/10.13039/501100011033, Plan National FIDEUA PID2019-106901GB-I00, FPI, QUANTERA MAQS PCI2019-111828-2, Proyectos de I+D+I “Retos Colaboración” QUSPIN RTC2019-007196-7); Fundació Cellex; Fundació Mir-Puig; Generalitat de Catalunya through the European Social Fund FEDER and CERCA program (AGAUR Grant No. 2017 SGR 134, QuantumCAT \ U16-011424, co-funded by ERDF Operational Program of Catalonia 2014-2020); EU Horizon 2020 FET-OPEN OPTOlogic (Grant No 899794); National Science Centre, Poland (Symfonia Grant No. 2016/20/W/ST4/00314); European Union’s Horizon 2020 research and innovation programme under the Marie-Sk{\l}odowska-Curie grant agreement No 101029393 (STREDCH) and No 847648 (“La Caixa” Junior Leaders fellowships ID100010434: LCF/BQ/PI19/11690013, LCF/BQ/PI20/11760031, LCF/BQ/PR20/11770012, LCF/BQ/PR21/11840013).

%\bibliography{ref_21}

%apsrev4-2.bst 2019-01-14 (MD) hand-edited version of apsrev4-1.bst
%Control: key (0)
%Control: author (8) initials jnrlst
%Control: editor formatted (1) identically to author
%Control: production of article title (0) allowed
%Control: page (0) single
%Control: year (1) truncated
%Control: production of eprint (0) enabled
%

%============================================ SUPPLEMENTAL MATERIAL ==================
\pagebreak
\section{Supplemental material}

%\date{\today}% It is always \today, today,

%  but any date may be explicitly specified

\maketitle

\setcounter{equation}{0}

\setcounter{figure}{0}

\setcounter{table}{0}

\makeatletter

\renewcommand{\theequation}{S\arabic{equation}}

\renewcommand{\thefigure}{S\arabic{figure}}

\renewcommand{\bibnumfmt}[1]{[S#1]}

\section{QVAE architecture}

\begin{figure}[h]
	\centering
	\includegraphics[width=\columnwidth]{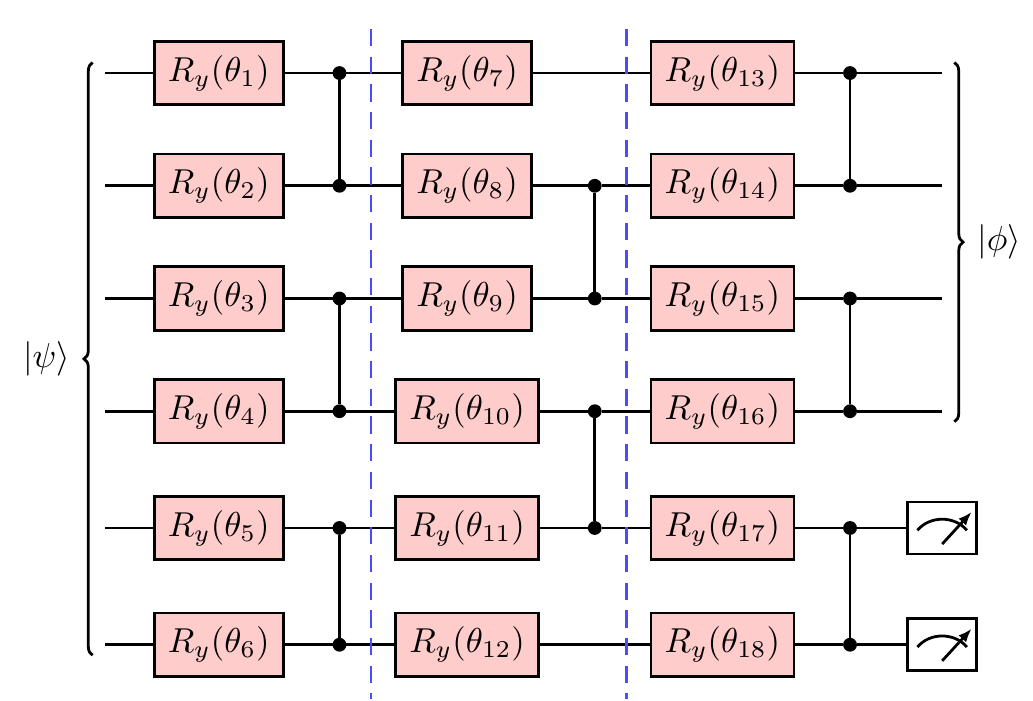}
	\caption{Quantum Variational Autoencoder (QVAE) composed of the single-qubit rotations around the $y$ axis parametrized by angles $\bm{\theta} = (\theta_1,\dots,\theta_{18})$ and two-qubit controlled-Z gates that entangle the neighboring qubits. Separate layers are denoted by dashed vertical lines.}
	\label{fig:circuit}
\end{figure}

The architecture of QVAE is an Alternating Layered Ansatz \cite{Cerezo21, Nakaji21} shown in Fig.~\ref{fig:circuit}. Optimal angles $\mathbf{\theta}$ that minimize cost defined in the main text are found by the SPSA optimizer \cite{Spall97, Spall98, Kandala17, Qiskit21short} with a random initial guess and automatic determination of the learning and perturbation rate implemented in Qiskit \cite{Qiskit21short}. In order to utilize the constraints imposed on the Hilbert space in the PXP model, we implement the unitary circuit $U(\bm{\theta})$ from scratch using the QuSpin Python library \cite{Weinberg17, Weinberg19}. The code is available upon request. %

Hyperparameters of the circuits: number of trash qubits $\Ntrash$, layers $L$, measurement shots in the determination of cost $\Nshots$ and training iterations $\Niter$ were found heuristically by performing a grid search for smaller PXP system with $N=18$ and choosing hyperparameters maximizing drops of cost on the $\Ztwo$ scars if trained on another $\Ztwo$ scar. Optimal parameters read $\Ntrash=6$, $L=5$, $\Nshots=300$, $\Niter=20000$, yet other sets with those quantities changed by factors of up to 25\% still yielded statistically significant drops of $\Ztwo$ scars costs. These results were the starting point for more computationally expensive $N=24$ case - they were scaled linearly with the system size and adjusted heuristically to optimal values $\Ntrash=8$, $L=7$, $\Nshots=600$, $\Niter=50000$ used to produce Figs.~1, 2 in the main text.

Inspection of the training outcomes can give intuition about the interplay of hyperparameters. It was observed that overfitting, i.e., a trivial learning of a perfect representation of the training eigenstate, which is easily detectable by a large drop in cost of the training point compared to all other eigenstates and no detection of other $\Ztwo$ scars in the benchmarking process, is caused by too many parameters (too large $L$) or too weak compression ratio (too small $\Ntrash$). The number of shots $\Nshots$ controls the variance of cost between iterations which should remain at around 1-10\% to overcome local minima but not jump too far in the cost landscape, in full analogy to the sizes of batches in the stochastic gradient descent algorithm \cite{Goodfellow16}. We also observed that optimal QVAE circuits had, unsurprisingly, more layers than CQVAE circuits for the same number of qubits, because of the need to represent an exponentially larger part of the Hilbert space with more parameters. We found that CQVAE was untrainable for the same number of layers as the QVAE. We expect this is due to the presence of four-qubit operations in CQVAE which makes the optimization problem more difficult. We also noticed that the number of layers $L$ should be large enough so that trash qubits are contained within ''light-cone'' of the first qubit.

\section{CQVAE operators}
Below we list exact matrix forms operators used in the QVAE variant restricted to constrained Hilbert space of the PXP model. Rotation of one qubit around the $y$ axis is performed only if its neighbours are in the state $\ket{0}$ according to 
\begin{equation}
	\tilde{R}_y(\theta) = \begin{blockarray}{ccccc}
		\ket{000}& \ket{010} & \ket{001} & \ket{100} & \ket{101} \\
		\begin{block}{(ccccc)}
			\cos (\theta/2) &  \sin(\theta/2) & & &   \\
			-\sin(\theta/2) & \cos(\theta/2) & & &  \\
			&  & 1 & & \\
			&  &  &1 &  \\
			&  &  & &1  \\
		\end{block}
	\end{blockarray}
\end{equation}
where states denoted above the matrix enumerate the 3-qubit computational basis with constraints. Similarly, two neighboring qubits are entangled only if their neighbors are in state $\ket{0}$ by the following operator:
\begin{equation}
	E_i = \begin{blockarray}{ccccccc}
		\ket{0000} & \ket{0010} & \ket{0100} & \ket{0001} & \ket{0101} & \ket{1000} & \ket{1010}  \\
		\begin{block}{(ccccccc)}
			1 &  &  & &  & &  \\
			&  1/\sqrt{2}& 1/\sqrt{2} & & & &  \\
			&  1/\sqrt{2}& -1/\sqrt{2} & & & &  \\
			&  &  & 1 &   &   &    \\
			&  &  &   & 1 &   &    \\
			&  &  &   &   & 1 &    \\
			&  &  &   &   &   & 1  \\
		\end{block}
	\end{blockarray}
\end{equation}
In this case, the entangling operator is different than in the $CZ$ gate (which acts not trivially only on 2-qubit configuration $\ket{11}$ excluded by the PXP model constraints).
\section{Detection of families in the PXP model}
\begin{figure*}
	\centering
	\includegraphics[width=\textwidth]{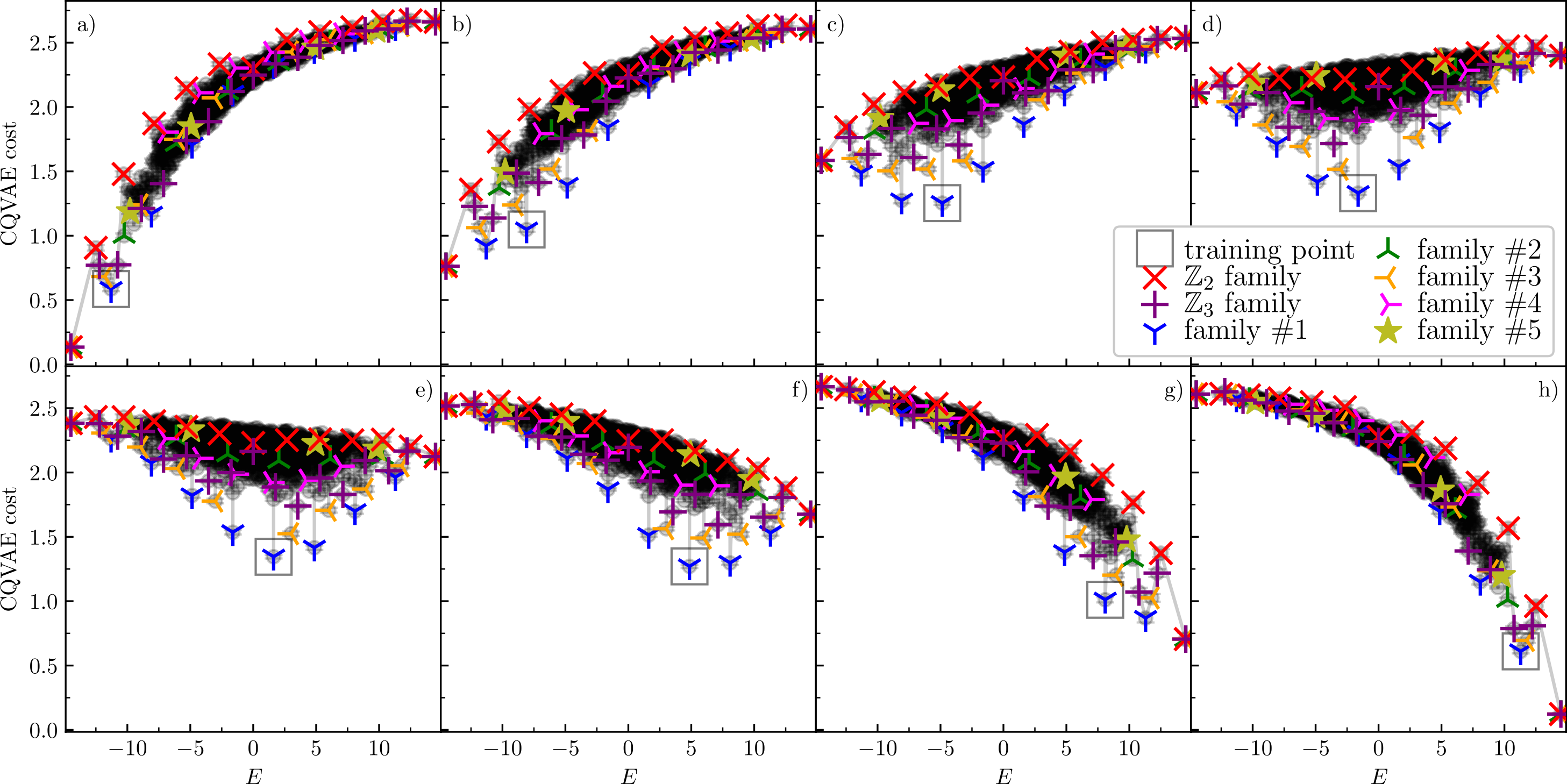}
	\caption{CQVAE cost vs energy for family $\#1$ for all eigenstates within this family. Above data were used to identify members of family $\#1$.}
	\label{fig:family_1}
\end{figure*}
We provide technical details concerning the scars identification in the PXP model. In Fig.~\ref{fig:family_1} we show output of CQVAE used to determine members of family $\#1$. We see that there is an overall growth of cost with distance in energy. Nevertheless, family members can be detected one by one using transitivity (see main text). Similar plots have been used to detect other families. Clusters of mutually related eigenstates (incl. transitivity) were identified by graph toolbox. Namely, we regard all eigenstates as graph nodes. For each training eigenstate, its corresponding node gets a directed link to nodes on which there are drops in QVAE cost. Then a community detection tool \texttt{CommunityGraphPlot} in Wolfram Mathematica is used to list all isolated families with more than one node and bidirectional links.

\section{Characterization of families in the PXP model}
In this paragraph we further characterize QMBS beyond the $\Ztwo$ and $\Zthree$ families detected by the QVAE in the PXP model. For families $\#1-\#5$ we notice an increased overlap with a set of Fock states listed in Table~\ref{tab:families}. For readability, Fock states are represented by $\Ztwo, \Zthree, \Zfour, \Zfive$ patterns that appear in them. For example, state denoted as $2-2-3-2-3-3-3-3-3$ corresponds to Fock state $\ket{101010010100100100100100}$. To illustrate the increased overlaps with Fock states from Table~\ref{tab:families}, we calculate probabilities of finding the eigenstates in any of those Fock states, 
\begin{equation}
	P(i) = \sum_{j=1}^{n_i} \left| \braket{E}{f_{ij}} \right|^2,
	\label{eq:prob}
\end{equation} where $\#i$ labels the family and $\lbrace\ket{f_{ij}}\rbrace_j$ are Fock states listed next to family $\#i$ in Table \ref{tab:families}, and plot them in Fig.~\ref{fig:overlap_Fock}.
\begin{table}
	\centering
	\begin{tabular}{lll}
		\toprule
		name & Fock state pattern & $(\#\Ztwo, \#\Zthree,\#\Zfour,\#\Zfive)$ \\
		\midrule
		\multirow{5}{*}{family \#1}  & 2-2-3-2-3-3-3-3-3     & (3,6,0,0)  \\
		& 2-3-2-3-2-3-3-3-3     & (3,6,0,0)  \\
		& 2-2-3-3-2-3-3-3-3     & (3,6,0,0)  \\
		& 2-3-2-3-3-2-3-3-3     & (3,6,0,0)  \\
		& 2-2-3-3-3-2-3-3-3     & (3,6,0,0)  \\
		\midrule
		\multirow{3}{*}{family \#2}  & 3-3-3-4-3-4-4         & (0,4,3,0)  \\
		& 3-3-4-3-3-4-4         & (0,4,3,0)  \\
		& 3-3-4-3-4-3-4         & (0,4,3,0)  \\
		\midrule
		family \#3                   & 2-2-2-3-2-3-2-2-3-3   & (6,4,0,0)  \\
		\midrule
		\multirow{5}{*}{family \#4}  & 2-2-2-2-2-2-2-2-2-3-3 & (9,2,0,0)  \\
		& 2-2-2-2-2-2-2-2-3-2-3 & (9,2,0,0)  \\
		& 2-2-2-2-2-2-2-3-2-2-3 & (9,2,0,0)  \\
		& 2-2-2-2-2-2-3-2-2-2-3 & (9,2,0,0)  \\
		& 2-2-2-2-2-3-2-2-2-2-3 & (9,2,0,0)  \\
		\midrule
		\multirow{10}{*}{family \#5} & 3-4-4-4-4-5           & (0,1,4,1)  \\
		& 3-4-4-4-5-4           & (0,1,4,1)  \\
		& 3-4-4-5-4-5           & (0,1,3,2)  \\
		& 3-4-4-5-4-4           & (0,1,4,1)  \\
		& 3-4-5-3-4-5           & (0,2,2,2)  \\
		& 3-4-5-3-5-4           & (0,2,2,2)  \\
		& 3-4-5-4-9             & (0,1,2,1)  \\
		& 3-4-5-4-3-5           & (0,2,2,2)  \\
		& 3-4-5-4-8             & (0,1,2,1)  \\
		& 3-4-9-3-5             & (0,2,1,1)  \\
		\bottomrule
	\end{tabular}
	\caption{List of Fock states with increased overlaps with corresponding QMBS families. Last column contains the number of occurences of corresponding patterns in the Fock states.}
	\label{tab:families}
\end{table}

We find that for families $\#1-\#4$ the numbers of patterns are conserved, and family $\#5$ gives more than one number of occurrences of each pattern. Family $\#4$ is composed of a maximal number of $\Ztwo$ patterns that can be supplemented by $\Zthree$ to fill the whole system with $N=24$. A similar family was observed for other system sizes: $N=18$ (6,2,0,0) and $N=30$ (12,2,0,0) at shifted energies, see Fig.~\ref{fig:PXP_N18_N24_N30}. It is at present not clear what is the mechanism behind the conservation of the ''domain wall'' number within a single family.

\section{Interpreting the QVAE circuit: input state optimization}
In the main text we demonstrate that QVAE allows one to find multiple scar families among eigenstates of the PXP model. In the paragraph above we listed physical features specific to the found families, indirectly indicating "what order parameter the QVAE learns". Here, we employ an alternative method to interpret the trained QVAE circuits. Variationally optimizing the input, starting from a random state, we simultaneously minimize the cost on trash qubits for an ensemble of independently trained QVAEs with constant circuit parameters. This approach is in stark contrast to standard training where the input is given and the circuit parameters are variationally adjusted. We optimize over an ensemble rather than a single QVAE for robustness of the result. Optimization of the input is a popular method of understanding layers in classical deep neural networks working with real-world pictures \cite{Erhan09, Yosinski15}, with a lot of optimization results especially entertaining to look at for humans \cite{Olah17}. Due to the generative nature of the process starting with random noise, input optimization is sometimes called "dreaming" and we adopt this naming here.

In FIG.~\ref{fig:dream}a) we show local magnetization $\langle \sigma_i^z \rangle$ in an example "dreamt" state $\ket{\chi_{dream}}$ found for ensemble of 100 CQVAEs trained on a $\Ztwo$ scar eigenstate in the middle of the spectrum. A $\Ztwo$-like magnetization pattern in $\ket{\chi_{dream}}$ can be noticed. It is also visible in the structure factor $S(q) = \sum_j \exp(i q j) \langle \sigma_j^z \rangle$ in FIG.~\ref{fig:dream}b) which has a peak at $q=\pi$. Therefore, even if we assume no prior knowledge that the $\ket{\Ztwo}$ state represents the $\Ztwo$ scar family, we are able to recover $\ket{\chi_{dream}} \approx \ket{\Ztwo}$, serving as a single representative of this family. Unfortunately, since the optimization is done in the input space, due to an exponential increase of computational time with system size, this method is not feasible for large systems. In FIG.~\ref{fig:dream} we used system size $N=12$, $50000$ input optimization steps, each with $100$ shots for cost estimation. Respective CQVAEs consisted of $L=2$ layers, $N_{\text{trash}}=4$ trash qubits, and were trained with $N_{\text{shots}}=100$ shots for cost estimation and $N_{\text{iter}}=1000$ parameter optimization steps.

\section{PXP-like model with a three-body blockade}
\begin{figure*}[ht]
	\raisebox{1cm}
	{
		%	\begin{figure}[h]
			%		\centering
			%		\includegraphics[width=.7\columnwidth]{dream.pdf}
			%		\caption{"Dreaming": optimizing the input state to minimize the QVAE cost (see text) simultaneously for 100 QVAEs trained on one $\Ztwo$ scarred eigenstate, $N=12$. We get a "representantive" state $\ket{\chi_{dream}} \approx \ket{\Ztwo}$ for the scar family. a) Local magnetization of the "dream" state $\ket{\chi_{dream}}$ shows a $\Ztwo$-like pattern. b) Structure factor of the calculated magnetization (averaged over 15 independent optimizations) has a peak at $q=\pi$, signaling a period-2 ordering in the position space.}
			%		\label{fig:dream}
			%	\end{figure}
		\begin{minipage}[b]{.99\columnwidth}
			\includegraphics[width=.9\columnwidth]{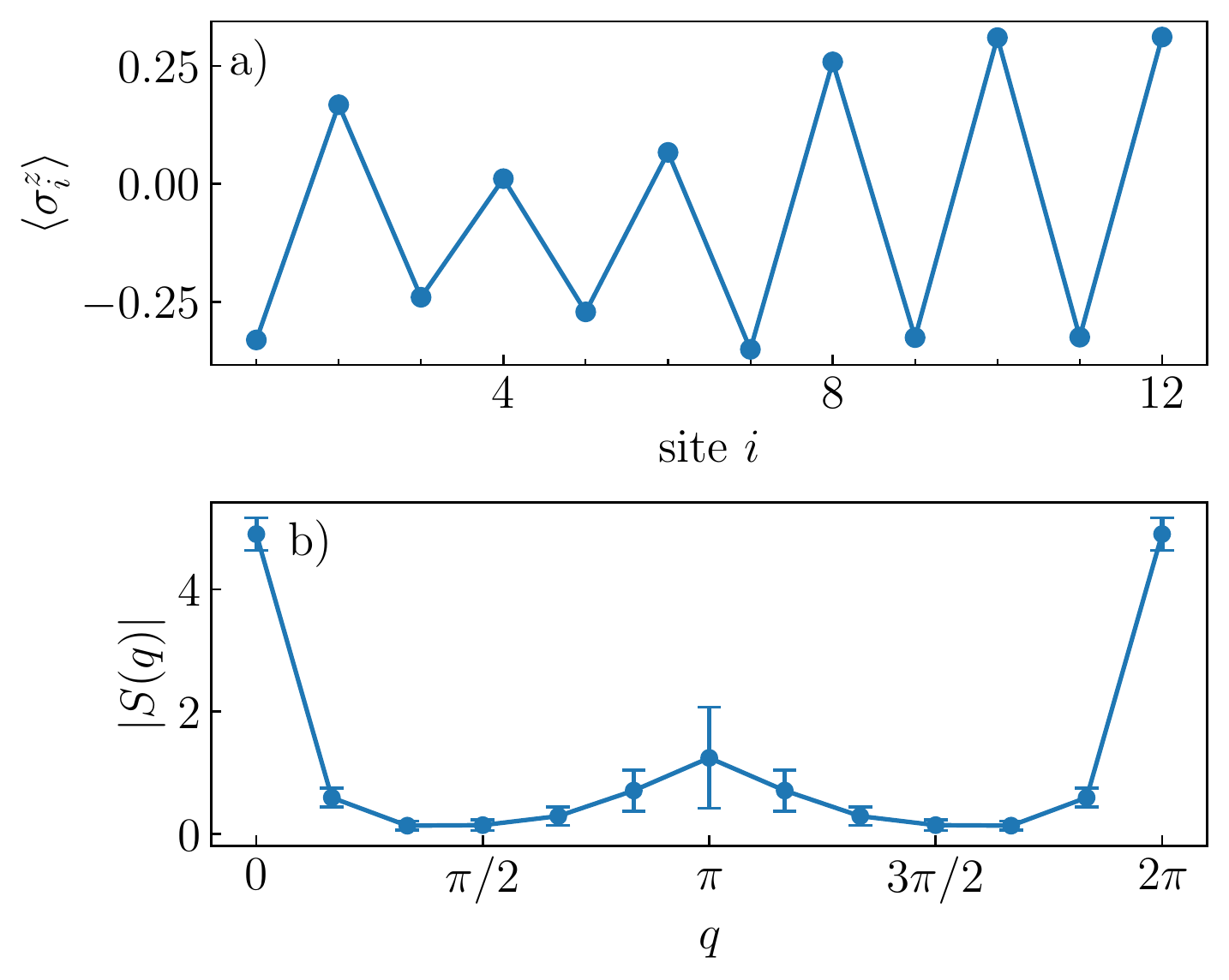}
			\caption{"Dreaming": optimizing the input state to minimize the QVAE cost (see text) simultaneously for 100 QVAEs trained on a single $\Ztwo$ scarred eigenstate, $N=12$. We get a "representantive" state $\ket{\chi_{dream}} \approx \ket{\Ztwo}$ for the scar family. a) Local magnetization of the "dream" state $\ket{\chi_{dream}}$ shows a $\Ztwo$-like pattern. b) Structure factor of the calculated magnetization (averaged over 15 independent optimizations) has a peak at $q=\pi$, signaling a period-2 ordering in the position space.}
			\label{fig:dream}
			%\vspace{2cm}
			%\end{minipage}
			%\begin{minipage}[b]{.99\columnwidth}
			\includegraphics[width=\columnwidth]{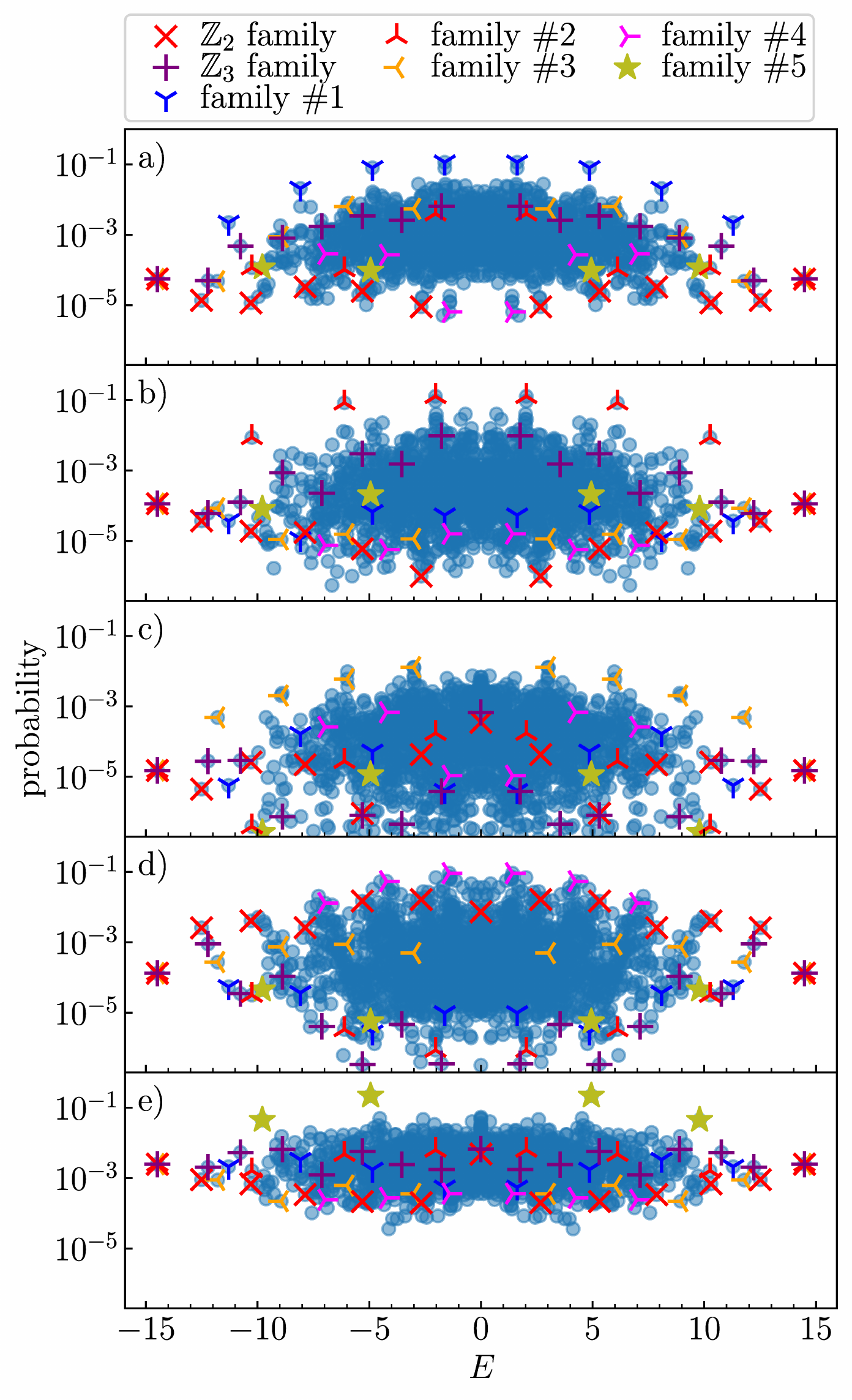}
			\caption{Probability (\ref{eq:prob}) of finding eigenstate in any of the Fock states the scarred family $\#1-\#5$ (a-e, respectively) have a large overlap with, see Table~\ref{tab:families}.}
			\label{fig:overlap_Fock}
		\end{minipage}
	}
	\hfill
	\raisebox{4cm}
	{
		\begin{minipage}[b]{.99\columnwidth}
			\includegraphics[width=\columnwidth]{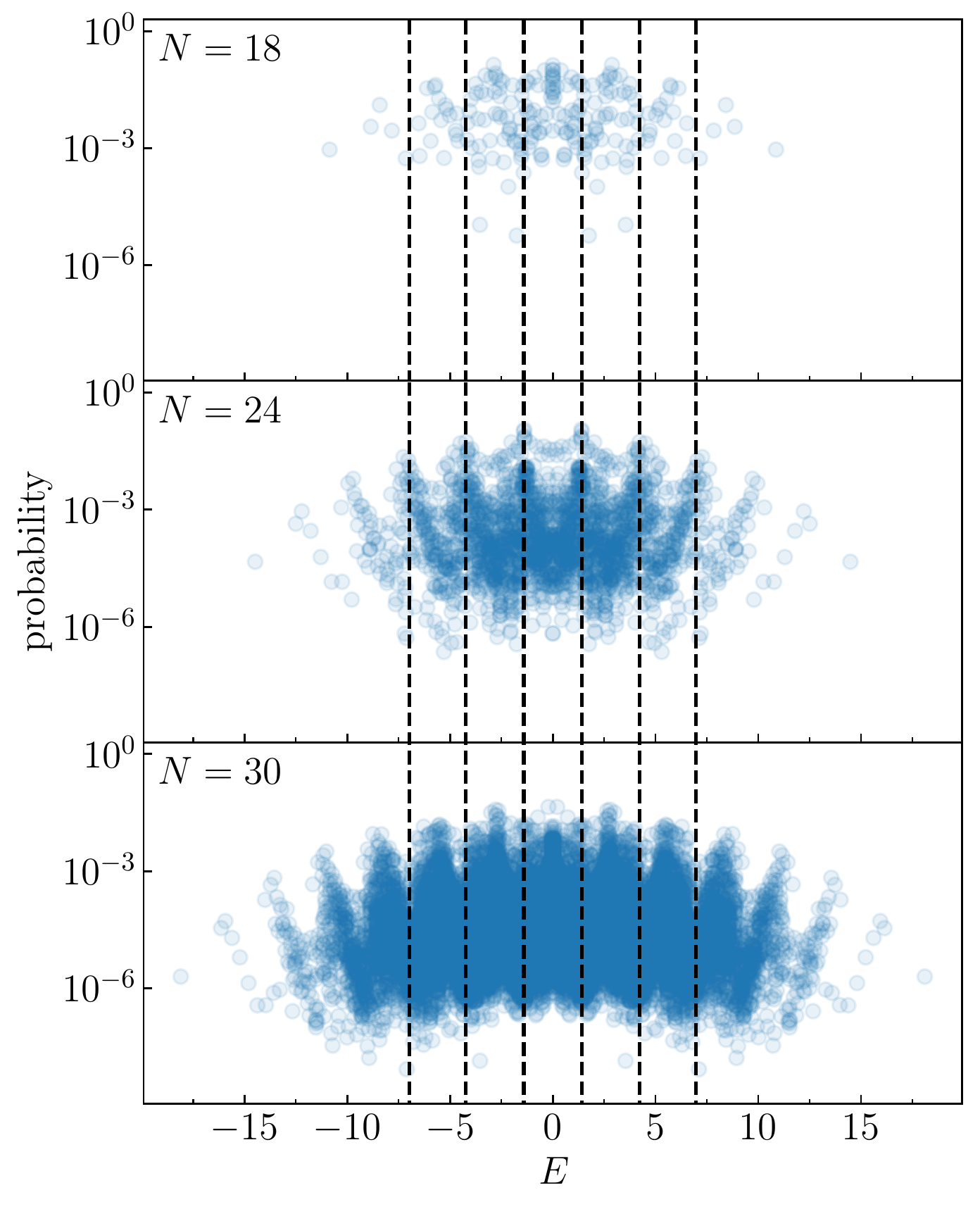}
			\caption{Probability (\ref{eq:prob}) of finding eigenstate in Fock states with (6,2), (9,2), (12,2) $\Ztwo$ and $\Zthree$ patterns for $N=18, 24, 30$, respectively. Energies of family $\#4$ found for $N=24$ by CQVAE are denoted by dashed vertical lines.}
			\label{fig:PXP_N18_N24_N30}
			\vspace{2cm}
			\includegraphics[width=\columnwidth]{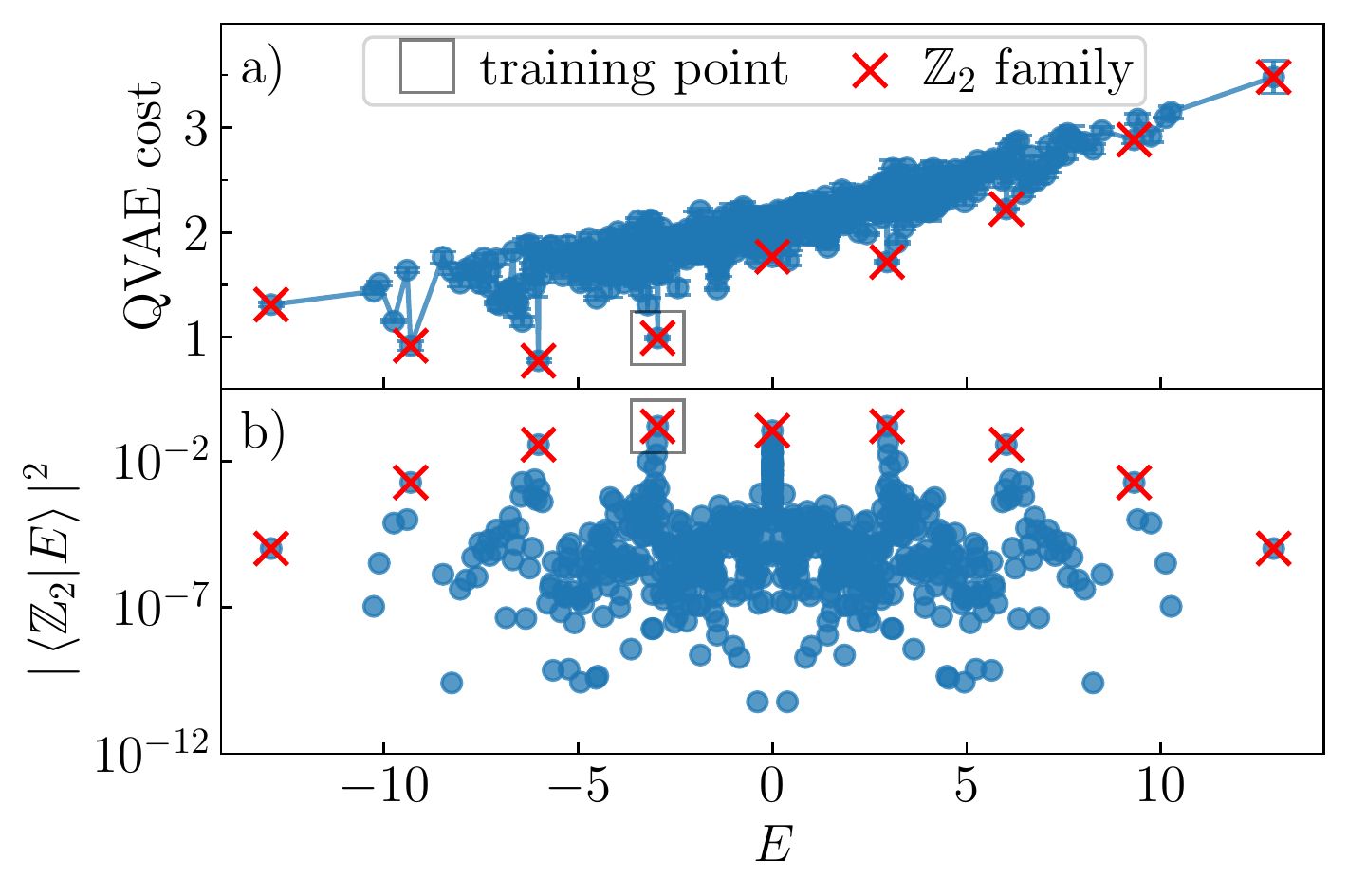}
			\caption{$\Ztwo$ scars detected by the QVAE in the variant of the PXP model with a three-body constraint.}
			\label{fig:qvae_pxp_3_body}
		\end{minipage}
	}
\end{figure*}

To demonstrate the universality of our method, we also apply it to a modified PXP model with another type of blockade \cite{Dooley21, Desaules21a}. The Hamiltonian reads
\begin{equation}
	\hat{H} = \sum_i \hat{P}_{i-1, i, i+1} \sigma_i^x,
\end{equation}
where $\hat{P}_{i-1, i, i+1}=1-\ket{1_{i-1}1_i 1_{i+1}}\bra{1_{i-1}1_i 1_{i+1}}$. This is a weaker constraint than in the PXP model. QVAE, working in the full Hilbert space, $\Ntrash=5$, $L=10$, applied to this model with $N=16$ detects scars of the $\Ztwo$ type, see Fig.~\ref{fig:qvae_pxp_3_body}. No other families have been found by the algorithm.
%\newpage
\end{document}